# *algoTRIC*: Symmetric and asymmetric encryption algorithms for Cryptography - A comparative analysis in AI era

Naresh Kshetri, Department of Cybersecurity, Rochester Institute of Technology, Rochester, New York, USA
Mir Mehedi Rahman, School of Business and Technology, Emporia State University, Emporia, Kansas, USA
Md Masud Rana, Department of Information Technology, San Juan College, Farmington, New Mexico, USA
Omar Faruq Osama, Dept. of System Science & Ind. Eng., Binghamton University, SUNY, New York, USA
James Hutson, Dept. of Art History, AI, & Visual Culture, Lindenwood University, St. Charles, Missouri, USA

**Abstract** - The increasing integration of artificial intelligence (AI) within cybersecurity has necessitated stronger encryption methods to ensure data security. This paper presents a comparative analysis of symmetric (SE) and asymmetric encryption (AE) algorithms, focusing on their role in securing sensitive information in AI-driven environments. Through an in-depth study of various encryption algorithms such as AES, RSA, and others, this research evaluates the efficiency, complexity, and security of these algorithms within modern cybersecurity frameworks. Utilizing both qualitative and quantitative analysis, this research explores the historical evolution of encryption algorithms and their growing relevance in AI applications. The comparison of SE and AE algorithms focuses on key factors such as processing speed, scalability, and security resilience in the face of evolving threats. Special attention is given to how these algorithms are integrated into AI systems and how they manage the challenges posed by large-scale data processing in multi-agent environments. Our results highlight that while SE algorithms demonstrate high-speed performance and lower computational demands, AE algorithms provide superior security, particularly in scenarios requiring enhanced encryption for AI-based networks. The paper concludes by addressing the security concerns that encryption algorithms must tackle in the age of AI and outlines future research directions aimed at enhancing encryption techniques for cybersecurity.

## I. Introduction

Algorithms are and were always the driving force behind cryptography and cybersecurity as we are marching towards the artificial intelligence (AI) and machine learning era. Several countermeasures, techniques, and cybersecurity practices are popular with the use of machine learning and deep learning algorithms apart from AI algorithms [1] [2]. As we know cybersecurity combines information security and network security, the annual number of data breaches is growing every year. Loss of private information, malware attacks, use of smart gadgets, growing number of internet population, and several others are upcoming challenges for cryptography algorithms.

Vulnerabilities and attacks on ciphers, private keys, and algorithms are increasing as we are considering "Security for AI" and vise-versa [3] [4]. New and unexpected attacks, development of several frameworks and tools are going on as we discuss various encryption algorithms. From the initial use of symmetric algorithms like Data Encryption Standard (DES), and their several weaknesses we tend to know that hackers are exploiting the powerful algorithms (like SHA3, MD5, up to CRYSTALS) today. The use of "secret key" in symmetric algorithms (although asymmetric works a little better as compared to symmetric) is no longer secret as attackers have successfully compromised the key in both symmetric and asymmetric algorithms.

---

The emergence of artificial intelligence (AI) has revolutionized cybersecurity, providing adaptive and dynamic encryption techniques to combat swiftly changing cyber threats. AI-driven methodologies have enhanced encryption systems' resilience, facilitating real-time identification of anomalies and threats that conventional methods find challenging to spot. The use of AI, especially via machine learning (ML) and deep learning (DL) algorithms, has markedly improved the efficacy of encryption methods, rendering them more adept at managing the increasing complexity and volume of contemporary data environments. [1] [2].

AI is becoming more and more integrated into cybersecurity and encryption as technology advances. AI is essential for protecting AI systems from complex cyberattacks, in addition to fortifying encryption procedures by streamlining key generation and data security techniques [Sec1Ref3]. In an increasingly linked and insecure digital world, the synergy between AI and encryption is essential because it allows more effective, scalable, and proactive security measures, guaranteeing the security of both data and AI systems [4].

This paper is structured as follows to explore the comparative analysis of encryption algorithms and their relevance in modern cybersecurity, particularly in the AI era. Section II provides a background study, outlining the historical evolution and challenges of cryptographic algorithms, and establishing the role of AI in enhancing these methods. Section III examines the technical aspects of various encryption algorithms, focusing on their contributions to safeguarding sensitive data in complex systems. Section IV offers a comparative analysis of symmetric encryption (SE) and asymmetric encryption (AE), highlighting key differences in terms of efficiency, security, and scalability. Section V discusses the role of AI in transforming encryption practices, focusing on how AI enhances real-time adaptability, tackles emerging threats, and enables personalized encryption strategies. Section VI focuses on the security challenges encryption faces in modern society, particularly against emerging cyber threats. Section VII presents the discussion and conclusions, summarizing the insights gained from the comparative analysis and suggesting improvements for existing encryption techniques. Lastly, Section VIII outlines the future scope of research, discussing potential advancements in encryption algorithms and their application in AI-driven cybersecurity solutions.

## II. Background Study

In [5], Kapoor and Thakur (2022) present a comprehensive comparative analysis of symmetric and asymmetric key algorithms, highlighting the escalating significance of encryption in safeguarding digital information within a progressively networked environment. Their analysis underscores the superiority of asymmetric key encryption over symmetric key encryption, as it employs two keys, hence augmenting security through mathematical complexity. The authors emphasize the adaptability and efficacy of AES as the preeminent symmetric algorithm, highlighting its resilience against prevalent assaults and rapid execution speed. Conversely, they recognize Elliptic Curve Cryptography (ECC) as the most secure asymmetric technique, chiefly due to its reliance on the algebraic framework of elliptic curves and finite fields. This technical examination of encryption algorithms is a vital reference for algoTRIC, particularly in assessing the performance of AES and ECC in AI-driven contexts. This paper examines the optimization of these algorithms inside algoTRIC's architecture for large-scale, multi-agent systems, emphasizing the critical trade-offs between speed and security resilience in addressing emerging cyber threats.

In [6], Soomro et al. (2019) conducted a comprehensive analysis of cryptographic algorithms, emphasizing both symmetric and asymmetric techniques and their contributions to improving cybersecurity across many contexts. Their paper delineates the fundamental cryptographic objectives, including secrecy, integrity, authenticity, and non-repudiation, which are vital for secure communications and data safeguarding across various sectors. The authors emphasize the importance of symmetric algorithms such as AES due to their speed and efficiency, rendering them suitable for extensive applications, while also addressing the strength of asymmetric algorithms, notably RSA, in contexts necessitating robust security and key management. This review is an essential resource for algoTRIC, providing fundamental insights regarding the adaptation of cryptographic approaches for the difficulties of AI-driven systems. Through the utilization of these encryption approaches in AI applications, algoTRIC examines the management of trade-offs among performance, scalability, and security resilience, especially in response to the escalating need for data protection within contemporary cybersecurity frameworks.

In [7], Ustun et al. (2021) introduce a sophisticated machine learning-based intrusion detection system aimed at mitigating cybersecurity vulnerabilities in smart grids using IEC 61850 Sampled Value (SV) messages. Their research highlights the significance of identifying cyberattacks, especially false data injection (FDI), inside contemporary power system communication frameworks, where both symmetric and asymmetric fault states are common. The proposed method utilizes machine learning to discriminate between regular grid operations and cyberattacks, demonstrating great accuracy in differentiating symmetrical faults, asymmetrical faults, and FDI attacks. This foundational study is essential for algoTRIC, as it highlights the potential of incorporating advanced intrusion detection algorithms into AI-driven environments, particularly if encryption techniques such as AES and ECC are utilized to secure communication streams. The findings of Ustun et al. offer significant insights into the management of data integrity and security by AI systems, which is essential for algoTRIC's investigation into reconciling encryption performance with the necessity for real-time intrusion detection in extensive, multi-agent contexts.

In [8], Arora (2022) presents a comprehensive analysis of cryptographic methods essential for cybersecurity, emphasizing the significance of encryption and decryption in safeguarding digital data. The study emphasizes the efficacy of symmetric encryption techniques, such as AES, for handling extensive data, as well as the strength of asymmetric algorithms like RSA for secure key management. Fundamental cryptographic principles, including confidentiality, integrity, and authenticity, are underscored as vital for safeguarding communication channels. This analysis provides essential insights for incorporating encryption techniques into AI-driven systems, where the equilibrium between computational performance and security is paramount. This analysis elucidates the trade-offs between the high performance of symmetric encryption and the superior security offered by asymmetric approaches, thereby guiding attempts to optimize encryption in systems that must navigate intricate cybersecurity difficulties while guaranteeing real-time data protection.

In [9], Henriques and Vernekar (2022) examine the amalgamation of symmetric and asymmetric cryptography to safeguard communication across devices in an IoT context. They emphasize the distinctive issues presented by IoT systems, wherein data transmitted among networked devices is extremely sensitive and requires protection against cyberattacks. Their methodology integrates the rapidity and efficacy of symmetric encryption with the robust key management of asymmetric cryptography to augment security. The implementation of AES for rapid encryption and RSA for secure key exchange mitigates prevalent IoT risks, including insecure network services and inadequate authentication procedures. This dual cryptographic methodology is pertinent to the design objectives of algoTRIC, since it provides an effective means of safeguarding communication in AI-driven, large-scale settings. This research utilizes both encryption algorithms to offer significant insights into balancing speed, scalability, and security in intricate multi-agent systems.

### III. Encryption Algorithms for Cybersecurity

Encryption techniques and complex algorithms with respect to privacy preserving, wireless sensor networks (WSN), and artificial intelligence (AI) are rapidly used in several system applications and solutions [10] [11]. Applications like healthcare monitoring, smart cities, advertising, logistics with analysis of energy, overhead, speed are used for several AI-powered business models. Financial transactions (may consist of hash, public key, private key, and digital signature) today require high level security using Secure Hashing Algorithms (SHA) and Message Digest (MD) algorithms used by distributed ledgers and blockchain technology.

Compressed sampling on encrypted images with a combined random Gaussian measurement matrix can also be used for AI based image encryption [12]. To resist several kinds of cyberattacks (primarily as primage attacks, collision attacks) that can pass plaintext sensitivity tests for successful communications. On the other hand, network security or endpoint security (of or partial of cryptography and/or cybersecurity), is fully achieved through data encryption using artificial intelligence [13]. Improving encryption speed, wireless sensors security, integrity of data proposed a proactive solution with remarkable performance as compared to static encryption methods.

Homomorphic encryption has arisen as a formidable method to bolster data security in AI-driven applications, facilitating computations on encrypted data without necessitating decryption. This capacity is essential for preserving data privacy in sensitive domains such as healthcare, banking, and smart city infrastructures, where AI is extensively employed for decision-making and data analysis. Homomorphic encryption encompasses several varieties, including fully homomorphic encryption (FHE), slightly homomorphic encryption (SWHE), and substantially homomorphic encryption (PHE), each presenting distinct trade-offs regarding computational complexity and efficiency [14]. Although Fully Homomorphic Encryption (FHE) permits infinite operations on encrypted data, its practical application is frequently constrained by substantial computing expenses and reduced processing velocities. Conversely, SWHE and PHE provide more efficient options by facilitating a limited range of actions, rendering them more appropriate for situations that emphasize performance while maintaining data security. In AI-driven contexts, including these encryption methods into machine learning models not only protects data during training and inference but also mitigates risks associated with emerging vulnerabilities such as data leakage and unauthorized access. As AI progresses, enhancing these encryption techniques will be essential for guaranteeing strong and scalable cybersecurity solutions.

The integration of AI approaches with conventional encryption algorithms such as AES has demonstrated effectiveness in augmenting data security, especially in volatile threat landscapes. Recent research indicates that the integration of machine learning models, such as k-Nearest Neighbors (k-NN), with AES encryption markedly enhances the identification and mitigation of anomalies, facilitating real-time responses to new cyber threats. The k-NN's pattern recognition capabilities enhance the encryption process, adapting to emerging attack vectors and bolstering AES's resilience against advanced attacks [15]. This method enhances secure data transmission and bolsters the integrity of secret data storage. With the increasing volume and complexity of data in AI-driven systems, integrating machine learning with encryption methods such as AES is crucial for adopting a proactive approach to cybersecurity.

Chaotic algorithms have arisen as an effective solution for image encryption in AI-driven networking systems, owing to its intrinsic characteristics such as sensitive dependence on beginning conditions, topological mixing, and long-term unpredictability [16]. These qualities are utilized to generate intricate encryption patterns, where even minor alterations in the original settings result in completely distinct encrypted outputs, hence substantially improving data security. Recent implementations indicate that chaotic algorithms, along with sophisticated encryption techniques, can provide non-linear transformations that effectively rearrange and disperse pixel positions, rendering the image data into a highly randomized state. This method guarantees that the encryption process emulates a dynamical system, rendering the reversal of the process without precise system parameters computationally impractical. Through the application of repeated chaotic functions, these encryption methodologies guarantee elevated entropy in the encrypted data, so successfully countering brute-force assaults and enhancing resilience against cryptographic scrutiny. In AI-driven environments, where data security is imperative against advanced threats, the amalgamation of chaotic systems with encryption enhances the security framework while preserving computational performance by reducing processing overhead [16].

Table 1. Intuitions (up to three) of some common advanced encryption algorithms for security and cryptography in the artificial intelligence (AI)-driven society

| Ref | Encryption Type | Intuition I | Intuition II | Intuition III |
| --- | --- | --- | --- | --- |

| [10] | Partial Homomorphic | Enables privacy-preserving computations on encrypted blockchain data | Mitigates risks from collision, preimage, and wallet attacks | Optimizes computational overhead for AI-integrated blockchain environments |
|---|---|---|---|---|
| [11] | AI-Driven Data Solutions | Adapts encryption parameters dynamically based on real-time network conditions | Integrates anomaly detection to proactively adjust encryption settings against threats | Optimizes computational and energy resources while maintaining high security levels |
| [12] | AI Image | Utilizes hyperchaotic sequences for robust pixel scrambling and diffusion | Enhances resistance against differential and brute-force attacks | Achieves high randomness and compression efficiency with compressed sensing |
| [13] | Innovative Data for WSANs | Adapts encryption parameters dynamically using AI for real-time threat response | Leverages LSTM networks to optimize encryption based on sequential data analysis | Employs Isolation Forests to enhance anomaly detection and network resilience |
| [14] | AI-based Homomorphic | Enables privacy-preserving computations on encrypted data without decryption | Mitigates data exposure risks in untrusted environments like cloud computing | Supports collaborative AI tasks with multi-key encryption across multiple parties |
| [15] | AI and AES | Combines AES's robust encryption with AI for adaptive threat detection | Utilizes AI-driven k-NN for real-time anomaly analysis in encrypted data | Enhances encryption efficiency through AI-optimized parameter selection |
| [16] | Image Transmission | Leverages chaotic mapping for high sensitivity and complex key generation | Enhances image confidentiality through pixel-level scrambling and diffusion | Mitigates brute-force attacks via topological chaos and statistical uniformity |

## IV. Comparisons of Algorithms w.r.t. SE and AE

To evaluate the cryptographic algorithms, it is significant to contrast symmetric encryption algorithms with asymmetric encryption algorithms as modern cryptography is designed based on symmetric and asymmetric encryption which are two fundamental categories of encryption algorithms. The main purposes of both types of encryption are the same, that is to safeguard the data security and integrity over the diverse applications. Although their purposes are the same, they have significant differences based on the way of managing encryption keys, evaluating performance and functionality requirements. To identify the most effective encryption method for a particular scenario, it is essential to distinguish the strength, weakness, functionalities, and other features of both types of encryption methods. This section of the paper distinguishes the fundamental types of encryption algorithm based on key management, scalability, swiftness, and reliability.

One of the main differences of symmetric and asymmetric encryption is the number of keys used in the encryption process. There are two types of keys used in encryption and decryption processes which are known as public key and private key. In a symmetric algorithm, a private key is used alone to encrypt and decrypt data. On the other hand, an asymmetric algorithm uses both the public key and private key where the public key is used to encrypt data and private key is used to decrypt data. Public key encryption is designed based on intensive computational mathematical functions; therefore, asymmetric algorithms are not very suitable or efficient for minor devices [17].

The second important term of differences between symmetric and asymmetric encryption is reliability. The encryption process of the symmetric method is simpler than the asymmetric method, however, in symmetric method both the sender and receiver share the common private key to encrypt and decrypt data which is a major concern about data security as eavesdropping can be conducted by attacker anytime in the channel of data exchange. Alternatively, in asymmetric encryption, the public key is used to encrypt the data while the private key is used to decrypt the data. As the private key is secret and only the receiver knows the private key, it becomes very difficult for the attacker to decrypt the original data. As a result, asymmetric encryption is considered more reliable in comparison to symmetric encryption in case of data exchange [18].

Swiftness of encryption and decryption is also a very powerful component that can be considered to differentiate symmetric and asymmetric encryption. Al-Shabi, in his paper, conducted an analysis to compare the performance for identifying the strengths and weaknesses of different types of symmetric and asymmetric encryption based on various factors such as battery consumption, block size, structure, time consumption and types of attacks. His result shows that based on real-time encryption, a symmetric algorithm is much faster than asymmetric encryption [18]. Similar kind of study was conducted by Panda in 2010. Her paper indicates that a symmetric algorithm is almost 1000 times faster than an asymmetric algorithm as an asymmetric algorithm needs more powerful computational resources. To compare different types of algorithm, 3 types of file such as text, image and binary were used in her analysis where the performance factors were decided considering Encryption Time, Decryption Time and Throughout. The result of her study found better performance from the AES algorithm, a subcategory of symmetric encryption, in comparison to other encryption algorithms based on Encryption Time, Decryption Time and Throughout [19].

Use of blocks is also a considerable component that can be used to distinguish between symmetric and asymmetric algorithms. There are mainly two important components considered in symmetric encryption known as block cipher and stream cipher, which are significantly crucial for confidentiality of data and integrity of cryptography [21]. AES, a subcategory of symmetric encryption, is operated on plaintext where the size of the block is 128 bits. This block cipher can also utilize different key lengths such as 128 bits, 192 bits or 256 bits of cipher secret [20]. On the other hand, asymmetric encryption does not require block size to encrypt data, rather this method leverages the idea of chunk data processing that is correspondent to the key size.

In the field of AI-driven cybersecurity, selecting between symmetric encryption (SE) and asymmetric encryption (AE) involves a thorough evaluation of performance, scalability, and security requirements. SE algorithms, such as AES, excel in real-time AI applications due to their high-speed encryption and low computational demands, which highlights as essential for AI tasks requiring rapid data processing [19]. However, AE algorithms like RSA provide enhanced security by leveraging public-private key pairs, a feature that underscores as crucial for maintaining confidentiality in sensitive data exchanges [18]. While SE is ideal for resource-constrained AI environments, such as IoT, due to its lower energy consumption, AE's computational intensity makes it better suited for secure initial key exchanges in distributed AI systems [20]. This difference in resource demands directly impacts scalability; SE supports continuous, high-throughput data streams often required in AI workflows, while AE's structure enables secure data sharing across complex, multi-agent networks through recent advances in secure communication protocols. [21] Effective cybersecurity in AI ultimately requires balancing SE's efficiency and AE's strong data protection, particularly in applications where threats to data integrity and confidentiality are significant [17].

Table 2: Comparison of Symmetric Encryption (SE) and Asymmetric Encryption (AE) in AI-Driven Cybersecurity

| Aspect | Symmetric Encryption (SE) | Asymmetric Encryption (AE) | Ref. |
|---|---|---|---|
| Integration with AI | SE algorithms like AES and Blowfish are efficient for real-time AI-driven data processing, supporting rapid encryption for high data volumes in AI workflows. | AE algorithms such as RSA and ECC are suitable for securely establishing initial connections in AI systems, though slower for real-time processing. | [19] |
| Data Throughput | High throughput makes SE ideal for handling large data in AI tasks (e.g., image processing or continuous data flows in AI-based IoT). | Lower throughput is better for secure, one-time exchanges rather than sustained high-speed AI-driven processing. | [21] |
| Resource Optimization | Low computational demands allow SE to support AI applications in resource-constrained environments, like mobile AI/IoT. | Higher resource needs make AE less suitable for low-power AI applications, though ideal for secure initial setup in complex AI networks. | [20] |
| Real-Time Efficiency | SE provides rapid encryption/decryption, enhancing real-time AI functions like anomaly detection in cybersecurity. | Slower speed limits AE in real-time AI scenarios; however, it provides robust security for secure data onboarding in AI systems. | [18] |
| Scalability in AI Systems | SE scales well within high-speed AI environments, enabling quick encryption across multi-agent or large data environments. | AE scales better for secure AI communications in distributed or cloud-based systems, especially for sensitive exchanges. | [21] |
| Battery and Power Use | Low power consumption suits AI-based mobile or IoT cybersecurity applications, allowing efficient continuous data encryption. | Higher power demand limits AE's suitability for battery-dependent AI devices, though it's viable for centralized secure key exchanges. | [20] |
| Security Strength | SE algorithms are faster but require secure key management in AI-driven environments to prevent compromise. | AE's public-private key pair provides greater security in AI-based networks with high confidentiality needs, particularly when securing data exchanges. | [18] |
| Complexity | Simpler structures in SE make it easier to embed into AI cybersecurity models needing rapid, low-latency responses. | AE's complexity is suitable for initial secure connections but can slow down ongoing high-volume AI data processing. | [19] |
| Use in AI Applications | Frequently applied in AI-driven real-time applications like intrusion detection, anomaly detection, and real-time threat monitoring. | Used to establish secure connections for sensitive AI operations, such as secure federated learning or distributed AI models. | [17] |
| Threat Resistance | Vulnerable to brute-force attacks if key length is inadequate; requires AI to monitor and update keys for robustness. | Robust against eavesdropping and man-in-the-middle attacks, providing strong security for AI-driven data exchanges across networks. | [21] |

## V. Algorithms in the Artificial Intelligence (AI) Era

AI is increasingly integrated into encryption techniques, offering adaptive and dynamic solutions to address evolving cybersecurity threats. ML models play a pivotal role by analyzing large datasets to detect anomalies, making encryption protocols more resilient to cyberattacks [22]. In recent years, there has been a surge in the application of deep learning to enhance cryptographic algorithms, particularly through convolutional neural networks (CNNs). These models help to create more robust key generation processes, as demonstrated in recent studies where CNNs were applied to Advanced Encryption Standard (AES) algorithms to improve encryption performance and security resilience [23]. Such AI-driven encryption systems are capable of continuously evolving, adapting to new security challenges, and countering sophisticated hacking attempts in real-time [24].

In addition to improving encryption processes, AI also aids in the proactive detection and mitigation of cyber threats. As Rangaraju [25] notes, through leveraging ML models, particularly deep learning algorithms, cybersecurity systems can predict potential vulnerabilities and strengthen encryption methods. These techniques not only enhance the overall security infrastructure but also allow for the development of intelligent, self-updating systems that can respond to newly emerging cyber threats. The real-time adaptability of AI in encryption is crucial, especially as traditional cryptography methods, such as RSA, become increasingly vulnerable to advanced cyberattacks [26]. This integration of AI into cryptography sets the stage for more secure communication and data protection in the AI era [27].

With the newly found ability to detect and mitigate cybersecurity threats, AI assists in offering advanced solutions that traditional encryption methods struggle to match. These solutions include CNNs, as noted, but also long short-term memory (LSTM), AI-driven systems that can analyze vast amounts of data in real-time, identifying patterns that signal potential threats. These AI-enhanced systems use data profiling techniques to categorize security events, enabling more accurate discrimination between legitimate threats and false positives [28]. For example, a study employing AI-based security information and event management (SIEM) demonstrated improved accuracy in detecting network intrusions by combining event profiling with various neural networks, outperforming traditional machine learning approaches [29] [30]. The ability to adapt to complex and evolving attack patterns makes these new technologies an essential tool for modern cybersecurity.

Such capacity to adapt and learn from emerging threats is critical as cybercriminals continuously develop more sophisticated attack methods. Deep learning models, especially when applied to real-time cybersecurity monitoring, can detect anomalies much faster than traditional methods, providing organizations with the agility to respond to cyberattacks proactively [31]. Recent advancements in deep learning-based intrusion detection systems (IDS) have shown promising results in identifying zero-day attacks, reducing detection time, and improving overall system security [32]. This proactive approach allows for not only quicker detection but also the anticipation of future attacks, helping organizations stay one step ahead of cybercriminals.

On the other hand, although integrating AI into encryption processes provides significant advancements and benefits, there are also numerous challenges and ethical concerns. One of the primary issues is the risk of over-reliance on AI-based systems, which could lead to complacency in monitoring and updating security protocols [33]. The dynamic nature of these tools can make encryption systems highly efficient, but this reliance also increases the risk that undetected vulnerabilities could be exploited by adversaries using AI for malicious purposes [34]. Furthermore, as AI-driven encryption systems become more widespread, the sheer volume of data processed raises concerns about privacy violations. AI models often require vast amounts of personal or sensitive information to function optimally, which can lead to unintended privacy breaches if not managed properly [35].

Another ethical concern involves the dual-use nature of AI technologies in encryption. While AI enhances security, it also opens avenues for adversaries to exploit AI systems to breach encrypted communications. AI-based algorithms could potentially be reverse-engineered or manipulated to bypass security protocols, creating a new type of cyber threat [36]. The sophistication of AI tools allows attackers to uncover hidden patterns or weaknesses in encryption systems, potentially leading to large-scale data breaches. This highlights the need for comprehensive governance frameworks that address not only the technical challenges but also the ethical risks associated with deploying AI in encryption and cybersecurity [37].

Looking ahead, ever-advancing AI tools are expected to play an increasingly central role in the future of encryption, evolving alongside the cyber threat landscape. The adaptability of AI to real-time data allows for personalized encryption solutions tailored to the behaviors and preferences of individuals, making it more difficult for cybercriminals to execute successful attacks [38]. Through learning from patterns in network traffic and user behavior, AI can continuously optimize encryption protocols, ensuring that they remain effective against emerging threats. This ability to adapt to new challenges positions the technology as a vital tool in maintaining robust cybersecurity defenses in the coming years [39].

Moreover, integration into encryption technologies opens possibilities for more seamless and efficient security solutions. The use of AI to automate encryption processes could lead to faster, real-time encryption adjustments without human intervention. This is particularly valuable in dynamic environments, such as the Internet of Things (IoT), where devices continuously communicate and exchange data [40]. The ability to monitor and respond to security threats in real-time ensures that encryption methods are always up to date, thus reducing the risk of breaches. However, these advancements must be balanced with considerations for ethical use and the prevention of potential misuse of AI in malicious hacking activities [41].

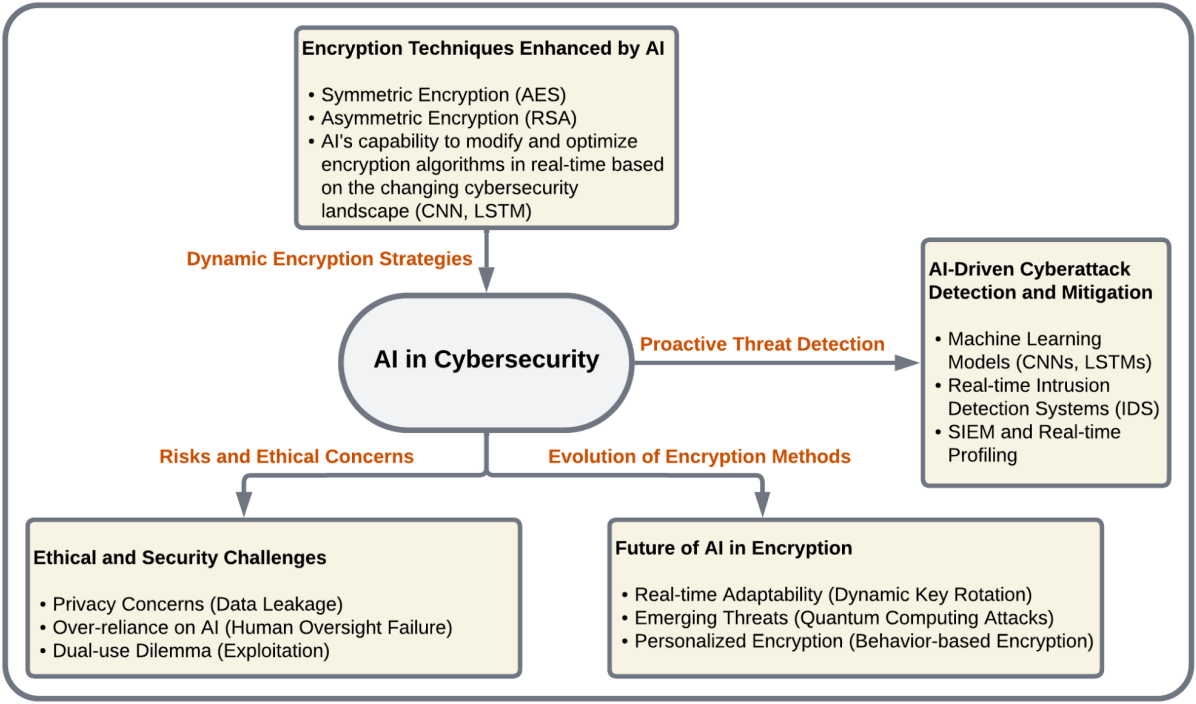

Figure 1: AI-Driven Enhancements in Encryption (including Symmetric and Asymmetric) and Cybersecurity

In the era of rapid technological progress, artificial intelligence has emerged as a revolutionary influence across several domains, including cybersecurity. As AI systems advance, the algorithms utilized for data protection as well as encryption must adapt to the intricacies of contemporary threats. The convergence of AI and encryption offers prospects for bolstering cybersecurity resilience via real-time monitoring, adaptive response strategies, and intelligent automation.

*5.1 Artificial Intelligence-Enhanced Encryption for Improved Cybersecurity*

Conventional encryption techniques, like Advanced Encryption Standard in symmetric encryption as well as RSA for asymmetric encryption, have recently been augmented using artificial intelligence to boost security and efficiency. The capacity of AI to analyze extensive data sets, identify trends, and adapt to emerging threats provides an ideal collaborator for cryptographic systems. AI-driven optimization strategies enhance symmetric encryption by dynamically creating and updating AES encryption keys according to real-time threat evaluations. Machine learning methods are now used to anticipate future weaknesses and thwart brute-force assaults by detecting anomalous patterns across encrypted data. This dynamic methodology is evolving AES into a more flexible and resilient system capable of adapting to various situations and threats without sacrificing speed [32].

On the asymmetric front, RSA encryption, which depends on both private and public keys, is benefiting from AI's capacity to improve the key generation process. Artificial intelligence methodologies, including genetic algorithms, have been employed to enhance the decision to use prime numbers, hence assuring that the produced encryption keys are more safe and less susceptible to assaults [23]. These improvements reduce the computing burden associated with both encryption and decryption, while preserving a robust degree of security. Moreover, deep learning methodologies, like convolutional neural networks (CNNs) along with recurrent neural networks (RNNs), are being included into cryptographic frameworks to oversee encrypted communications in real-time. These algorithms can detect irregularities in encrypted data streams, identify prospective breaches, and react preemptively to prevent system intrusion [33]. This real-time detection, driven by AI, offers an extra degree of protection that static encryption technologies alone cannot deliver. The use of AI in encryption enhances security systems while also improving their speed and efficiency. Incorporating convolutional neural networks (CNNs) within the AES key generation operations has been shown to improve security while decreasing computing cost [28]. Asymmetric cryptographic methods, such as RSA, may use AI approaches to enhance the efficiency of encryption and decryption operations in resource-limited situations such as IoT [29].

## *5.2 Homomorphic Encryption and Privacy-Enhancing Artificial Intelligence*

One of the most exciting advancements in AI-driven encryption involves the progression of homomorphic encryption. Homomorphic encryption enables calculations upon encrypted data without necessitating decryption, so safeguarding sensitive information during processing. This is especially beneficial in AI applications requiring the analysis of large data sets, such as in finance, healthcare as well as cloud computing.

Artificial intelligence is significantly enhancing the efficiency and scalability of homomorphic encryption techniques. Utilizing AI methodologies might enhance the efficacy of homomorphic encryption by reducing the noise typically accumulated during calculations, hence making these techniques more appropriate for practical use [35]. This advancement is particularly significant for privacy-preserving AI applications, in which sensitive data, such as health-related records and financial information, must be safeguarded throughout the analytical process [22]. Homomorphic encryption, in conjunction with AI, allows businesses to cooperate upon encrypted data without disclosing the underlying knowledge. This privacy-preserving methodology has considerable ramifications for sectors such as healthcare, where patient information may be safely exchanged and evaluated across institutions without jeopardizing privacy or regulatory adherence [39].

## *5.3 Blockchain and Artificial Intelligence: A Collaborative Strategy for Security*

Blockchain technology, recognized because of its decentralized and safe characteristics, is further enhanced using AI. The integration of AI into blockchain systems is yielding novel encryption methods that improve the security as well as effectiveness of blockchain transactions. AI-driven homomorphic encryption alongside other sophisticated cryptographic methods are used to guarantee that data sent over blockchain networks stays protected and safe.

AI is enhancing consensus techniques such as proof-of-work (PoW) along with proof-of-stake (PoS) inside blockchain systems. These AI-augmented algorithms raise the velocity and diminish the energy expenditure of blockchain networks, hence improving efficiency without compromising security [40]. Moreover, AI is used to identify and alleviate security problems in blockchain networks instantaneously, reinforcing blockchain's status as a fundamental element of safe digital transactions [37].

*5.4 Artificial Intelligence and Quantum-Resistant Cryptography*

As quantum computing progresses, the encryption techniques safeguarding our digital communications encounter considerable problems. Quantum computers, capable of resolving intricate mathematical problems far more rapidly than classical computers, present a risk to conventional cryptographic methods such as RSA along with elliptic curve cryptography (ECC). AI is being used to create quantum-resistant encryption methods.

Artificial intelligence methodologies are used to develop and evaluate post-quantum cryptographic algorithms that withstand assaults from quantum computers. A possible method is lattice-based cryptography, which depends on the complexity of resolving lattice issues, a challenge that persists regardless of quantum computers [30]. Artificial intelligence enhances these algorithms by examining possible weaknesses and assuring their effective deployment in practical systems [27]. Furthermore, AI is used to forecast the evolution of quantum computers and to ensure that encryption methods remain resilient against these advancements. Through the simulation of quantum assaults and the evaluation of encryption systems, AI is contributing to the development of more robust cryptographic standards designed to safeguard sensitive information in the quantum age [26].

*5.5 Ethical Implications in AI-Enhanced Cryptography*

The incorporation of AI within encryption systems presents significant ethical dilemmas. As AI algorithms increase in complexity, the need for openness and accountability in their decision-making processes, especially in encryption and cybersecurity, is intensifying. It is essential to design AI-driven cryptography systems with ethical concerns to foster confidence and avoid abuse. A primary worry is the dual-use characteristic of AI technology. Although AI may improve encryption as well as cybersecurity, it may also be utilized by nefarious individuals to develop more advanced assaults or to avoid detection. Developing AI-driven encryption systems with strong ethical standards is crucial to avoid their misuse for bad reasons [36]. Furthermore, as AI along with encryption technologies proliferate, it is essential to guarantee their accessibility and equity. It includes tackling the digital divide including guaranteeing that modern encryption technologies are accessible to all societal sectors, not just to those with the means to use them [24].

To get farther into the AI age, encryption algorithms must advance to match the increasing sophistication of cyber threats. Artificial intelligence is significantly transforming both asymmetric and symmetrical encryption systems, which renders them more adaptable, effective, and safe. The integration of AI in key generation and real-time threat detection is transforming cybersecurity methodologies. Nonetheless, the prospect of AI-driven cryptography has concerns as well. It is essential for these systems to be morally robust, transparent, and resilient against new dangers, including those from quantum computing, to ensure their success. Advancing and perfecting AI-driven encryption methods will enable the establishment of an increased secure digital future which safeguards sensitive information while promoting innovation.

## VI. Algorithm Security in Modern Society

Encryption algorithms are essential tools in maintaining the confidentiality and integrity of digital communications in modern society. With the increasing reliance on digital platforms for both personal and professional interactions, ensuring secure communication has become a priority [42]. Algorithms such as the AES and RSA are widely adopted to protect sensitive data, including emails, financial transactions, and other online communications. AES, a symmetric key algorithm, is favored for its speed and efficiency in encrypting large volumes of data, making it suitable for applications where rapid data processing is critical [43]. In contrast, RSA, an asymmetric key algorithm, is often used for secure key exchanges and digital signatures due to its robust security features, although it operates at a slower speed [44]. Together, these algorithms form the foundation of secure digital communications, providing the first line of defense against unauthorized access and cyberattacks.

As society becomes more dependent on digital communication, the application of encryption algorithms continues to expand. For instance, hybrid encryption schemes that combine the strengths of both AES and RSA are becoming more popular. These hybrid systems leverage the efficiency of AES in data encryption and the strength of RSA in secure key management, ensuring that both the data and the encryption keys are protected during transmission [45]. Such combined approaches offer enhanced security, particularly in environments where large volumes of sensitive information are frequently exchanged, such as in e-commerce or financial institutions [46]. As encryption technologies evolve, they continue to play a vital role in safeguarding digital communication, adapting to new threats and ensuring that sensitive information remains confidential and secure [47]. Thus, actionable risk assessment methodologies are particularly valuable for organizations that rely heavily on algorithms for their security, as they provide a clear framework to assess vulnerabilities, adapt to evolving threats, and reduce reliance on external vendors [48].

Yet, as noted, the rapid adoption of IoT devices and cloud computing has created new vulnerabilities in cybersecurity systems, particularly due to the limited computing capabilities of many IoT devices [34]. Many of these devices rely on lightweight encryption algorithms, such as the Data Encryption Standard (DES) or AES, which are efficient but may be more susceptible to attacks due to their reduced complexity [49]. Additionally, IoT devices often lack regular security updates, making them easy targets for cybercriminals. Cloud computing environments further complicate the situation, as data in transit and at rest in the cloud are vulnerable to interception, especially during migration between different cloud platforms [49]. This growing complexity necessitates the development of more robust encryption techniques tailored to the needs of both IoT and cloud environments [51].

Furthermore, the rise of supply chain attacks, where third-party software or hardware components are compromised, presents another significant challenge. As Hammi Zeadally and Nebhen (2023) point out, since many organizations rely on cloud services that integrate multiple external vendors, ensuring the security of every component is increasingly difficult [52]. In such environments, traditional encryption methods may not provide sufficient protection against sophisticated attacks. Emerging encryption models, such as lattice-based cryptography and hybrid encryption schemes, have been proposed as solutions to strengthen security, especially in resource-constrained IoT devices and cloud platforms [53]. As IoT and cloud ecosystems continue to expand, the demand for advanced encryption methods that can effectively address these new vulnerabilities will only increase [54]. Also, The escalating sophistication of cryptojacking and ransomware highlights the importance of robust encryption algorithms to safeguard against unauthorized access and financial disruptions who are using blockchain technology for their security [55].

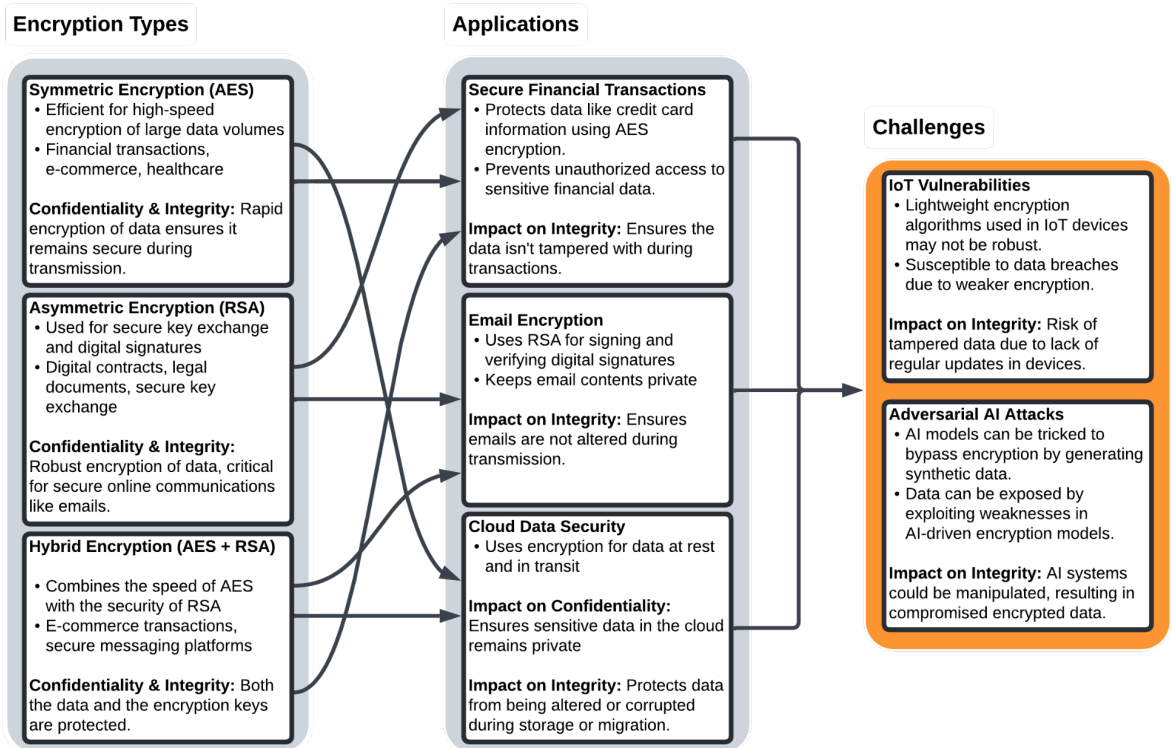

Figure 2: Encryption Algorithms (Symmetric encryption, Asymmetric encryption, and Hybrid encryption) Securing Digital Communications and Challenges

While these tools can enhance encryption and cybersecurity, it also introduces new vulnerabilities, particularly through adversarial AI attacks. These attacks exploit the weaknesses in AI models by introducing adversarial inputs, causing the system to make incorrect decisions. In the context of encryption, adversaries can manipulate these models designed to detect anomalies in encrypted communications or tamper with ML algorithms that generate encryption keys [56]. For example, recent studies have shown that adversarial ML techniques can be used to bypass AI-driven encryption models by generating synthetic data that mimics normal traffic patterns, thereby fooling detection systems [57].

Moreover, adversarial attacks can target not just encryption algorithms but the entire AI-based cybersecurity framework. These attacks can render AI-based defenses ineffective by exploiting weaknesses in neural networks used for real-time threat detection [58]. For instance, Generative Adversarial Networks (GANs) have been employed to create realistic attack scenarios that deceive AI systems, making it harder for traditional encryption methods to safeguard data [59]. The increasing sophistication of adversarial AI raises the stakes for maintaining secure systems, requiring not only advancements in encryption but also in AI model robustness [60]. As these threats evolve, the integration of more secure AI models into encryption protocols will be vital for protecting sensitive information in the digital age.

Moreover, the widespread use of encryption technologies in sectors such as finance, healthcare, and national security brings with it significant ethical and legal challenges. Governments and regulatory bodies face the difficult task of balancing individual privacy rights with the need for surveillance to prevent criminal activities [61]. Encryption ensures that sensitive data remains confidential, but it also makes it harder for law enforcement agencies to access potentially crucial information [63]. As a result, there has been ongoing debate about the implementation of encryption backdoors, which would allow authorized entities to decrypt data under specific circumstances. However, these backdoors present a serious ethical dilemma, as they could be exploited by malicious actors if not properly secured [63]. As encryption continues to play a critical role in modern society, it will be essential for policymakers to develop clear, globally consistent frameworks that address both the ethical and legal challenges posed by these technologies [37]. In addition to the ethical concerns, encryption technologies also raise legal questions regarding jurisdiction and data ownership. As data crosses international borders, determining which country's laws apply to encrypted information becomes increasingly complicated [64]. For instance, different nations have varying regulations regarding data privacy and encryption standards, which can lead to conflicts when encrypted data is stored in one country but accessed or processed in another [65].

## VII. Discussion and Algorithm Issues

The integration of artificial intelligence (AI) with encryption signifies a pivotal change in cybersecurity, offering both prospects and complex obstacles. The significance of AI in encryption has led to significant progress constantly in real-time threat detection as well as flexible security mechanisms, which are more vital in the contemporary linked and susceptible digital environment. This capacity allows encryption systems to promptly address abnormalities and emerging attack patterns, hence providing resilience unattainable by conventional static encryption approaches. Nonetheless, this progress entails an increasing dependence on machine learning as well as deep learning models, that, whilst augmenting encryption capabilities, can present weaknesses like adversarial assaults. These assaults target vulnerabilities in AI models using misleading inputs, compromising the precision and resilience of systems intended to identify and counter cyber threats. Thus, the dual-use characteristic of AI technology requires a measured and attentive strategy, especially in vital sectors such as healthcare, banking, and national security, wherein AI-driven encryption plays a crucial role in safeguarding extremely sensitive information.

Furthermore, the widespread use of AI-driven encryption systems raises urgent accessibility and ethical issues. Even while these technologies provide scalable solutions, their use in a variety of international businesses raises concerns about transparency and equality, particularly when firms have varying levels of technological resources as well as regulatory compliance skills. The fair distribution of these cutting-edge technologies must be given equal weight with technological resilience in the advancement of AI-powered encryption. These technologies also raise significant ethical and legal issues, including surveillance, data privacy, and the possible abuse of AI-enhanced encryption to provide vulnerabilities for illegal data access. To prevent AI-encrypted systems from unintentionally jeopardizing the same security and confidentiality they are meant to safeguard, strict governance structures and ethical standards must be established. Therefore, this open conversation covers both the enormous possibilities and the serious threats of AI-driven encryption, necessitating a thorough, interdisciplinary response to responsibly influence cybersecurity's future.

To sum up, the combination of encryption and artificial intelligence has brought about a new age in cybersecurity that offers increased resistance to a wide range of online dangers. AI-driven encryption is essential in today's fast-paced, data-intensive digital environment because of its adaptable, real-time features, which provide major benefits over conventional encryption methods. Homomorphic encryption and AI-enhanced algorithms are only two examples of the encryption techniques that have advanced because of this integration, strengthening data security, and enabling sophisticated calculations on encrypted data. AI algorithms provide a strong defense against complex cyberattacks as they become better at managing the complexities of threat detection including adaptive encryption key management. However, this development raises fresh moral and legal issues. The ethical conundrums around privacy, transparency, and equality, together with the dual-purpose possibilities for AI technology, highlight the need for a concerted effort from all parties involved. To create moral guidelines including legal frameworks that encompass both the technical aspects of AI-enhanced encryption and its wider social ramifications, cooperation between government, business, and academia is crucial.

In the future, establishing a safe and flexible cybersecurity framework will require a proactive approach to the creation and management of AI-driven encryption systems. To reduce new dangers and protect sensitive data in a variety of industries, further research in fields like adversarial resilience, quantum-resistant encryption, and ethical AI will be essential. Through adopting this forward-thinking viewpoint, the cybersecurity industry can capitalize on AI's ability to develop encryption technologies while additionally making certain that those solutions are just, ethically appropriate, as well as resilient to the constantly changing cyberthreat scenario. In this sense, incorporating AI into encryption seems not just a technological development but also a step toward a digital future that is safe, sustainable, as well as considerate of privacy.

Although there are several issues in algorithms for both encryption and decryption, some of the major ones (in symmetric encryption and asymmetric encryption) are shown in Figure 3 below. The challenges in algorithm generation, algorithm writing, and algorithm difficulty continues as the use of various language models including Artificial Intelligence (AI), Deep Learning (DL), and Machine Learning (ML) keeps growing.

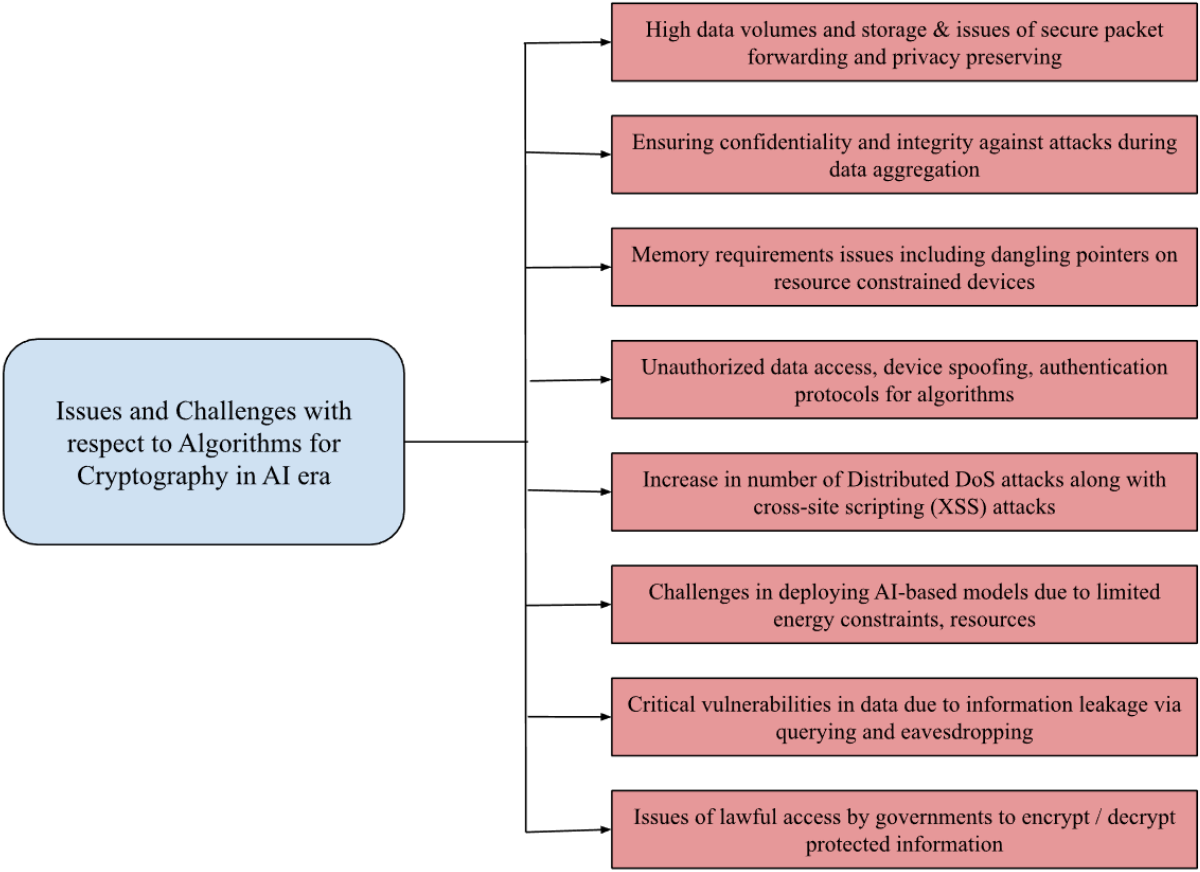

Figure 3: Encryption Algorithms Issues and Challenges for Cryptography / Cybersecurity in AI era (including both Symmetric, Asymmetric algorithms) [51] [53] [54] [57] [58] [60] [62]

## VIII. Conclusion and Future Scope

We provided an in-depth study focusing on securing sensitive information with comparative analysis for symmetric encryption and asymmetric encryption algorithms for cryptography. The comparison on the study focuses on key factors like security resilience, scalability, speed in the light of evolving cyber threats. We have addressed the security concerns tackled by encryption algorithms in the Artificial Intelligence (AI) and Large Language Models (LLMs) age along with research directions to enhance overall cybersecurity and cryptography. The aspect comparison between symmetric encryption and asymmetric comparison allows us to decide the environments used including AI environments, key-pairs leveraging via secure key exchanges, and/or decision in secure protocols for secure data sharing.

Future scope of algorithms whether it be symmetric encryption and asymmetric encryption algorithms, largely rely upon use of AI models, reliability, scalability, and key management. Enabling machines to learn is always a future challenge that may require human intelligence in the next step for decision-making. As we progress into several AI algorithms in the future, all three types of learning (supervised learning, unsupervised learning, and reinforcement learning) we must be more intuitive in the future on how we process data and information.

References
[1] Thiyagarajan, P. (2020). A review on cyber security mechanisms using machine and deep learning algorithms. *Handbook of research on machine and deep learning applications for cyber security*, 23-41.
[2] Terumalasetti, S., & Reeja, S. R. (2022, August). A comprehensive study on review of AI techniques to provide security in the digital world. In *2022 third international conference on intelligent computing instrumentation and control technologies (ICICICT)* (pp. 407-416). IEEE.
[3] Al-Arjan, A., Rasmi, M., & AlZu'bi, S. (2021, July). Intelligent security in the era of AI: The key vulnerability of RC4 algorithm. In *2021 International Conference on Information Technology (ICIT)* (pp. 691-694). IEEE.
[4] Bertino, E., Kantarcioglu, M., Akcora, C. G., Samtani, S., Mittal, S., & Gupta, M. (2021, April). AI for Security and Security for AI. In *Proceedings of the Eleventh ACM Conference on Data and Application Security and Privacy* (pp. 333-334).
[5] Kapoor, J., & Thakur, D. (2022). Analysis of symmetric and asymmetric key algorithms. In *ICT analysis and applications* (pp. 133-143). Springer Singapore.
[6] Soomro, S., Belgaum, M. R., Alansari, Z., & Jain, R. (2019, August). Review and open issues of cryptographic algorithms in cyber security. In *2019 International Conference on Computing, Electronics & Communications Engineering (iCCECE)* (pp. 158-162). IEEE.
[7] Ustun, T. S., Hussain, S. S., Yavuz, L., & Onen, A. (2021). Artificial intelligence based intrusion detection system for IEC 61850 sampled values under symmetric and asymmetric faults. *Ieee Access*, *9*, 56486-56495.
[8] Arora, S. (2022). A review on various methods of cryptography for cyber security. *Journal of Algebraic Statistics*, *13*(3), 5016-5024.
[9] Henriques, M. S., & Vernekar, N. K. (2017, May). Using symmetric and asymmetric cryptography to secure communication between devices in IoT. In *2017 International Conference on IoT and Application (ICIOT)* (pp. 1-4). IEEE.
[10] Yaji, S., Bangera, K., & Neelima, B. (2018, December). Privacy preserving in blockchain based on partial homomorphic encryption system for AI applications. In *2018 IEEE 25th International Conference on High Performance Computing Workshops (HiPCW)* (pp. 81-85). IEEE.
[11] Arulmurugan, L., Thakur, S., Dayana, R., Thenappan, S., Nagesh, B., & Sri, R. K. (2024, May). Advancing Security: Exploring AI-driven Data Encryption Solutions for Wireless Sensor Networks. In *2024 International Conference on Advances in Computing, Communication and Applied Informatics (ACCAI)* (pp. 1-6). IEEE.
[12] Xu, D., Li, G., Xu, W., & Wei, C. (2023). Design of artificial intelligence image encryption algorithm based on hyperchaos. *Ain Shams Engineering Journal*, *14*(3), 101891.
[13] Dharmateja, M., Rama, P. K., Asha, N., Nithya, P., Lalitha, S., & Manojkumar, P. (2024, March). Innovative Data Encryption Techniques using AI for Wireless Sensor Actuator Network Security. In *2024 International Conference on Distributed Computing and Optimization Techniques (ICDCOT)* (pp. 1-6). IEEE.
[14] Hamza, R. (2023, October). Homomorphic Encryption for AI-Based Applications: Challenges and Opportunities. In *2023 15th International Conference on Knowledge and Systems Engineering (KSE)* (pp. 1-6). IEEE.
[15] Budhewar, A., Bhumgara, S., Tekavade, A., Nandkar, J., & Zanwar, A. (2024, April). Enhancing Data Security through the Synergy of AI and AES Encryption: A Comprehensive Study and Implementation. In *2024 MIT Art, Design and Technology School of Computing International Conference (MITADTSoCiCon)* (pp. 1-5). IEEE.
[16] Tian, H., Yuan, Z., Zhou, J., & He, R. (2024). Application of Image Security Transmission Encryption Algorithm Based on Chaos Algorithm in Networking Systems of Artificial Intelligence. In *Image Processing, Electronics and Computers* (pp. 21-31). IOS Press.
[17] Abd Elminaam, D. S., Abdual-Kader, H. M., & Hadhoud, M. M. (2010). Evaluating the performance of symmetric encryption algorithms. *Int. J. Netw. Secur.*, *10*(3), 216-222.
[18] Al-Shabi, M. A. (2019). A survey on symmetric and asymmetric cryptography algorithms in information security. *International Journal of Scientific and Research Publications (IJSRP)*, *9*(3), 576-589.
[19] Panda, M. (2016, October). Performance analysis of encryption algorithms for security. In *2016 International Conference on Signal Processing, Communication, Power, and Embedded System (SCOPES)* (pp. 278-284). IEEE.
[20] Hintaw, A. J., Manickam, S., Karuppayah, S., Aladaileh, M. A., Aboalmaaly, M. F., & Laghari, S. U. A. (2023). A robust security scheme based on enhanced symmetric algorithm for MQTT in the Internet of Things. *IEEE Access*, *11*, 43019-43040.
[21] Kuznetsov, O., Poluyanenko, N., Frontoni, E., & Kandiy, S. (2024). Enhancing Smart Communication Security: A Novel Cost Function for Efficient S-Box Generation in Symmetric Key Cryptography. *Cryptography*, *8*(2), 17.
[22] Halewa, A. S. (2024). Encrypted AI for Cyber security Threat Detection. *International Journal of Research and Review Techniques, 3*(1), 104-111.

[23] Negabi, I., El Asri, S. A., El Adib, S., & Raissouni, N. (2023). Convolutional neural network based key generation for security of data through encryption with advanced encryption standard. *International Journal of Electrical & Computer Engineering* (2088-8708), 13(3).
[24] Rehan, H. (2024). AI-Driven Cloud Security: The Future of Safeguarding Sensitive Data in the Digital Age. *Journal of Artificial Intelligence General science (JAIGS)* ISSN: 3006-4023, 1(1), 132-151.
[25] Rangaraju, S. (2023). Ai sentry: Reinventing cybersecurity through intelligent threat detection. *EPH-International Journal of Science And Engineering, 9*(3), 30-35.
[26] Saha, A., Pathak, C., & Saha, S. (2021). A Study of Machine Learning Techniques in Cryptography for Cybersecurity. *American Journal of Electronics & Communication, 1*(4), 22-26.
[27] Yanamala, A. K. Y., & Suryadevara, S. (2023). Advances in Data Protection and Artificial Intelligence: Trends and Challenges. *International Journal of Advanced Engineering Technologies and Innovations, 1*(01), 294-319.
[28] Feisheng, L. (2024, April). Systematic Review of Sentiment Analysis: Insights Through CNN-LSTM Networks. In *2024 5th International Conference on Industrial Engineering and Artificial Intelligence (IEAI)* (pp. 102-109). IEEE.
[29] Pacheco, J., Benitez, V. H., Felix-Herran, L. C., & Satam, P. (2020). Artificial neural networks-based intrusion detection system for internet of things fog nodes. *IEEE Access, 8*, 73907-73918.
[30] Tashfeen, M. T. A. (2024). Intrusion detection system using AI and machine learning algorithms. In *Cyber security for next-generation computing technologies* (pp. 120-140). CRC Press.
[31] Mallick, M. A. I., & Nath, R. (2024). Navigating the Cyber security Landscape: A Comprehensive Review of Cyber-Attacks, Emerging Trends, and Recent Developments. *World Scientific News, 190*(1), 1-69.
[32] Alionsi, D. D. D. (2023). AI-driven cybersecurity: Utilizing machine learning and deep learning techniques for real-time threat detection, analysis, and mitigation in complex IT networks. *Advances in Engineering Innovation, 3*, 27-31.
[33] Orner, C., & Chowdhury, M. M. (2024). AI and Cybersecurity: Collaborator or Confrontation. *Proceedings of 39th International Confer, 98*, 150-158.
[34] Jimmy, F. N. U. (2024). Cyber security Vulnerabilities and Remediation Through Cloud Security Tools. *Journal of Artificial Intelligence General science (JAIGS)* ISSN: 3006-4023, 2(1), 129-171.
[35] Gupta, A., Wright, C., Ganapini, M. B., Sweidan, M., & Butalid, R. (2022). State of AI ethics report (volume 6, february 2022). *arXiv preprint arXiv:2202.07435*.
[36] Riebe, T. (2023). Dual-Use and Trustworthy? A Mixed Methods Analysis of AI Diffusion between Civilian and Defense R&D. In *Technology Assessment of Dual-Use ICTs: How to Assess Diffusion, Governance and Design* (pp. 93-110). Wiesbaden: Springer Fachmedien Wiesbaden.
[37] Gupta, M., Akiri, C., Aryal, K., Parker, E., & Praharaj, L. (2023). From chatgpt to threatgpt: Impact of generative ai in cybersecurity and privacy. *IEEE Access*.
[38] Evren, R., & Milson, S. (2024). The Cyber Threat Landscape: Understanding and Mitigating Risks. *Tech. rep. EasyChair*.
[39] Morley, J., & Floridi, L. (2020). An ethically mindful approach to AI for health care. *The Lancet, 395*(10220), 254-255.
[40] Javadpour, A., Ja'fari, F., Taleb, T., Zhao, Y., Bin, Y., & Benzaïd, C. (2023). Encryption as a service for IoT: opportunities, challenges, and solutions. *IEEE Internet of Things Journal*.
[41] Gupta, A., Royer, A., Heath, V., Wright, C., Lanteigne, C., Cohen, A., Ganapini, M., Fancy, M., Galinkin, E., Khurana, R., Akif, M., Butalid, R., Khan, F., Sweidan, M., Institute, A., , M., Cambridge, U., Commons, C., Exeter, U., University, C., Lab, A., Global, A., , M., College, U., Toronto, U., Ottawa, U., , R., NYUCenterforResponsible, A., Hyderabad, I., & University, M. (2020). The State of AI Ethics Report (October 2020). *ArXiv, abs/2011.02787*.
[42] Thabit, F., Can, O., Aljahdali, A. O., Al-Gaphari, G. H., & Alkhzaimi, H. A. (2023). Cryptography algorithms for enhancing IoT security. *Internet of Things, 22*, 100759.
[43] Kuppuswamy, P., Al, S. Q. Y. A. K., John, R., Haseebuddin, M., & Meeran, A. A. S. (2023). A hybrid encryption system for communication and financial transactions using RSA and a novel symmetric key algorithm. *Bulletin of Electrical Engineering and Informatics, 12*(2), 1148-1158.
[44] Pandey, P. K., Kansal, V., & Swaroop, A. (2023). Security challenges and solutions for next-generation VANETs: an exploratory study. In *Role of Data-Intensive Distributed Computing Systems in Designing Data Solutions* (pp. 183-201). Cham: Springer International Publishing.
[45] Akter, R. I. M. A., Khan, M. A. R., Rahman, F. A. R. D. O. W. S. I., Soheli, S. J., & Suha, N. J. (2023). RSA and AES based hybrid encryption technique for enhancing data security in cloud computing. Int. J. Comput. *Appl. Math. Comput. Sci, 3*, 60-71.
[46] Liu, Y., Gong, W., & Fan, W. (2018). Application of AES and RSA Hybrid Algorithm in E-mail. *2018 IEEE/ACIS 17th International Conference on Computer and Information Science (ICIS)*, 701-703. https://doi.org/10.1109/ICIS.2018.8466380.
[47] Subramanian, A., Donta, L. S., & Supraja, P. (2024, May). Assessing the Strength of Hybrid Cryptographic Algorithms: A Comparative Study. In *2024 International Conference on Intelligent Systems for Cybersecurity (ISCS)* (pp. 1-6). IEEE.
[48] Rahman, M. M., Kshetri, N., Sayeed, S. A., & Rana, M. M. (2024). AssessITS: Integrating procedural guidelines and practical evaluation metrics for organizational IT and cybersecurity risk assessment. *Journal of Information Security, 15*(4), 564–588. https://doi.org/10.4236/jis.2024.154032
[49] Singh, S., Sharma, P. K., Moon, S. Y., & Park, J. H. (2024). Advanced lightweight encryption algorithms for IoT devices: survey, challenges, and solutions. *Journal of Ambient Intelligence and Humanized Computing,* 1-18.
[50] Siwakoti, Y. R., Bhurtel, M., Rawat, D. B., Oest, A., & Johnson, R. C. (2023). Advances in IoT security: Vulnerabilities, enabled criminal services, attacks, and countermeasures. *IEEE Internet of Things Journal, 10*(13), 11224-11239.
[51] Zhou, J., Cao, Z., Dong, X., & Vasilakos, A. (2017). Security and Privacy for Cloud-Based IoT: Challenges. *IEEE Communications Magazine, 55*, 26-33. https://doi.org/10.1109/MCOM.2017.1600363CM
[52] Hammi, B., Zeadally, S., & Nebhen, J. (2023). Security threats, countermeasures, and challenges of digital supply chains. *ACM Computing Surveys, 55*(14s), 1-40.
[53] Bagla, P., Sharma, R., Mishra, A., Tripathi, N., Dumka, A., & Pandey, N. (2023). An Efficient Security Solution for IoT and Cloud Security Using Lattice-Based Cryptography. *2023 International Conference on Emerging Trends in Networks and Computer Communications (ETNCC)*, 82-87. https://doi.org/10.1109/ETNCC59188.2023.10284931.
[54] Ahmed, S., & Khan, M. (2023). Securing the Internet of Things (IoT): A comprehensive study on the intersection of cybersecurity, privacy, and connectivity in the IoT ecosystem. *AI, IoT and the Fourth Industrial Revolution Review, 13*(9), 1-17.
[55] Kshetri, N., Rahman, M. M., Sayeed, S. A., & Sultana, I. (2024). *cryptoRAN: A review on cryptojacking and ransomware attacks w.r.t. banking industry - Threats, challenges, & problems.* In *2024 2nd International Conference on Advancement in Computation & Computer Technologies (InCACCT)* (pp. 523–528). IEEE. https://doi.org/10.1109/InCACCT61598.2024.10550970

[56] Craighero, F., Angaroni, F., Stella, F., Damiani, C., Antoniotti, M., & Graudenzi, A. (2023). Unity is strength: Improving the detection of adversarial examples with ensemble approaches. *Neurocomputing, 554,* 126576.

[57] Shroff, J., Walambe, R., Singh, S. K., & Kotecha, K. (2022). Enhanced security against volumetric DDoS attacks using adversarial machine learning. *Wireless Communications and Mobile Computing, 2022*(1), 5757164.

[58] Sathupadi, K. (2023). Ai-based intrusion detection and ddos mitigation in fog computing: Addressing security threats in decentralized systems. *Sage Science Review of Applied Machine Learning, 6*(11), 44-58.

[59] Zhang, C., Yu, S., Tian, Z., & Yu, J. J. (2023). Generative adversarial networks: A survey on attack and defense perspective. *ACM Computing Surveys, 56*(4), 1-35.

[60] Fernando, P., & Wei-Kocsis, J. (2021). A Novel Data Encryption Method Inspired by Adversarial Attacks. *ArXiv, abs/2109.06634.*

[61] Allahrakha, N. (2023). Balancing cyber-security and privacy: legal and ethical considerations in the digital age. *Legal Issues in the Digital Age,* (2), 78-121.

[62] van Daalen, O. L. (2023). The right to encryption: Privacy as preventing unlawful access. *Computer Law & Security Review, 49,* 105804.

[63] Taddeo, M., McCutcheon, T., & Floridi, L. (2019). Trusting artificial intelligence in cybersecurity is a double-edged sword. *Nature Machine Intelligence,* 1-4. https://doi.org/10.1038/s42256-019-0109-1.

[64] Lubin, A. (2023). The prohibition on extraterritorial enforcement jurisdiction in the datasphere. In *Research Handbook on Extraterritoriality in International Law* (pp. 339-355). Edward Elgar Publishing.

[65] Nguyen, M. T., & Tran, M. Q. (2023). Balancing security and privacy in the digital age: an in-depth analysis of legal and regulatory frameworks impacting cybersecurity practices. *International Journal of Intelligent Automation and Computing, 6*(5), 1-12.